\documentclass[%
aip,
 amsmath,amssymb,
 reprint,%
]{revtex4-1}

\usepackage{graphicx}
\usepackage{dcolumn}
\usepackage{bm}

\usepackage[utf8]{inputenc}
\usepackage[T1]{fontenc}
\usepackage{mathptmx}
\usepackage{etoolbox}
\usepackage[ruled,vlined]{algorithm2e}
\usepackage{siunitx}

\makeatletter
\def\@email#1#2{%
 \endgroup
 \patchcmd{\titleblock@produce}
  {\frontmatter@RRAPformat}
  {\frontmatter@RRAPformat{\produce@RRAP{*#1\href{mailto:#2}{#2}}}\frontmatter@RRAPformat}
  {}{}
}%
\makeatother
\begin{document}

\preprint{AIP/123-QED}

\title[Sample title]{The Principle of Minimum Pressure Gradient: An Alternative Basis for Physics-Informed Learning of Incompressible Fluid Mechanics}

\author{H. Alhussein}
\affiliation{ 
Engineering Division, New York University Abu Dhabi, Abu Dhabi, UAE
}

\author{M. Daqaq}
\email{haa385@nyu.edu}
\affiliation{ 
Engineering Division, New York University Abu Dhabi, Abu Dhabi, UAE
}
\affiliation{ 
Tandon School of Engineering, New York University, Brooklyn, NY, USA
}

\date{\today}

\begin{abstract}
Recent advances in the application of physics-informed learning into the field of fluid mechanics have been predominantly grounded in the Newtonian framework, primarly leveraging  Navier-Stokes Equation or one of its various derivative to train a neural network. Here, we propose an alternative approach based on variational methods. The proposed approach uses the principle of minimum pressure gradient combined with the continuity constraint to train a neural network and predict the flow field in incompressible fluids. We describe the underlying principles of the proposed approach, then use a demonstrative example to illustrate its implementation and show that it reduces the computational time per training epoch when compared to the conventional approach.  
\end{abstract}

\maketitle

\section{\label{sec:Intro}Introduction}
Before digital computing, the field of mechanics relied mostly on analytical tools, which placed limitations on our ability to solve many important problems that are described by complex multi-dimensional and/or nonlinear partial differential equations (PDEs). To tackle them, even the most astute mathematicians need to invoke simplifying assumptions often diluting the underlying physics.   Motivated by the need to solve such problems, and enabled by the availability of fast computing resources, the field of computational mechanics grew rapidly over the past five decades becoming the method of choice for delineating the complex mechanics of solids, fluids, and their interactions \citep{hughes2000finite, belytschko2000nonlinear,ferziger2002computational, dowell2001modeling}. To this end, various computational methods including finite element (FE) and finite volume (FV) methods, as well as spectral techniques were developed and employed  \citep{belytschko2000nonlinear, versteeg2007introduction, boyd2001chebyshev}. 

Despite such advancements, traditional computational techniques are continually challenged by several enduring complexities. Mesh generation remains a time-consuming process that is an art form for most complex domains involving interacting energy fields. Moreover, solving inverse computational problems, such as those involving unknown boundary conditions is often prohibitively expensive. Commercial software packages such as COMSOL, ANSYS, and OpenFOAM involve extensive amount of coding at the backend to handle solver algorithms, mesh generation, and to set up the boundary conditions. This makes the process of maintenance and generational updates an astounding task.

Amid these enduring challenges, the field of computational mechanics has recently turned towards unconventional solutions hoping to overcome these hurdles. One such promising avenue is the application of machine learning techniques to solve complex PDEs \citep{lagaris1998artificial}.  However, largely due to the absence of advanced computational tools such as efficient deep machine learning algorithms and the advancements in GPU technology, such approaches did not gain significant traction until 2019, when Raissi et al. \citep{raissi2019physics} introduced the concept of physics-informed deep learning (PINN).  In principle, the proposed method starts by using a neural network to initialize a random guess for the unknown dependent variables, then substitutes them into the governing PDEs, initial, and boundary conditions. A loss function is defined and used to measure the error between the predicted output based on the guessed variables and the target solution. The loss function is then minimized with respect to the parameters of the neural network. Iterations continue until the loss function converges to a predefined threshold around zero. 

The development of PINN has inspired a number of applications in fluid mechanics involving the prediction of flows fields in various domains \citep{cai2021physics,eivazi2022physics,li2022machine,kissas2020machine,wang2021deep,wang2020towards}. The main attractive advantage of PINNs is that a unified flexible and robust framework can be used for both forward and inverse problems \citep{chen2020physics, gao2022physics}. Moreover, compared to traditional computational fluid dynamics (CFD) solvers, PINNs are superior at incorporating real-world observations of the flow quantities into the governing equations. A notable example is presented in Ref. \citep{cai2021Heat}, where PINNs were proposed to infer the coefficient of the convection term in the Navier-Stokes equations (NSE) based on experimental velocity measurements for the 2D flow over a cylinder.  

In the domain of fluid mechanics, the application of PINNs is predominantly grounded in the Newtonian framework, primarily leveraging NSE or one of its various derivatives \citep{gresho1987pressure}. Unlike in other domains where variational methods have effectively solved numerous problems \citep{lanczos2012variational}, the use of these principles in fluid mechanics has been somewhat limited. This limitation stems from the less mature theoretical foundation in fluid mechanics. For instance, extensive studies have attempted to extend the formulation of the NSE to align with Hamilton's principle of least action. However, these efforts have faced obstacles, largely because the Hamiltonian framework struggles to incorporate non-conservative forces like viscosity \citep{bretherton1970note, salmon1988hamiltonian, morrison1998hamiltonian}.  As a result, variational principles in fluid mechanics have faced constraints in effectively addressing the complexities of real-world scenarios.

In a departure from traditional approaches, Taha and Gonzales \citep{gonzalez2022variational, taha2023minimization} introduced a transformative concept in fluid mechanics based on Gauss' principle of least constraint, leading to the development of the Principle of Minimum Pressure Gradient (PMPG). This principle asserts that, within any incompressible flow, the pressure gradient is minimized at all times. This framework diverges markedly from conventional methodologies, offering a unique variational approach to the problems of fluid dynamics. 

Building upon this foundation, our research introduces a paradigm shift in addressing incompressible flow problems by combining PMPG with PINN, thus creating PMPG-PINN. This approach uniquely informs the neural network's loss function by focusing on minimizing a cost function derived from the PMPG that is independent of the pressure. The removal of the pressure dependence from the unknown variables reduces computational cost both in terms of reducing the size of the unknowns, and in terms of evaluating derivatives in the loss function. To the authors' knowledge, this is the first work that combines PMPG and PINN to solve incompressible flow problems.

To elucidate our proposed method, the paper is structured as follows. Section 2 provides a mathematical formulation of the principle of PMPG, shedding light into its theoretical foundation. Section 3 illustrates the conventional implementation of PINN using Navier-Stokes Equations. Section 4 demonstrates the newly proposed implementation of PINN using PMPG. Section 5 presents an illustrative example that showcases the effectiveness of the approach in solving incompressible flow problems. Finally, Section 6 presents the conclusions, and potential applications and extensions of the proposed method.

\section{Principle of Minimum Pressure Gradient}\label{sec:FP}
To effectively present the PMPG, a foundational understanding of Gauss' principle of least constraint is essential. This principle, which forms the core of the derivation, is extensively detailed in references \citep{udwadia2002foundations, udwadia2023general}. Consider the dynamics of $N$  constrained particles, each of a fixed mass $m_i$, whose motion can be described by the generalized coordinates, $\mathbf{q}$.  The dynamics of such particles is dictated by Newton's second law as: 
\begin{equation}
    m_i\; a_i(\Ddot{q}, \dot{q}, q) = F_i+R_i, \qquad i=1,2, \ldots N,
\end{equation}
where $a_i$ is the inertial acceleration of the $i$th particle, $F_i$ are the external forces acting on the particle, and $R_i$ are the constraint forces, which do zero work as they preserve the constraints. 

Gauss' principle asserts that the quantity 
\begin{equation}
    S = \frac{1}{2} \sum_{i=1}^{N} m_i \left( a_i - \frac{F_i}{m_i}\right)^2
\end{equation}
is a minimum with respect to the generalized accelerations, $\Ddot{\mathbf{q}}$, at every instant of time.  An equivalent $S$ can be written in terms of the constraint forces as:  
\begin{equation}
    S = \frac{1}{2} \sum_{i=1}^{N}  \left(\frac{R_i}{m_i}\right)^2,
\end{equation}
where the sum of the squares of the constraint force must be a minimum. 

In the absence of constraint forces, $R_i = 0$, $S$ assumes its absolute minimum, which is equal to zero. In other words, a particle follows the external force applied to it. However, in a constrained setting, nature acts like a mathematician. It picks, at each instant of time, an acceleration for the constrained system that minimizes, in a weighted least-squares sense, the difference between the acceleration of the free motion and the constrained system. This means that, the particle adjusts its path only to the extent necessary to meet the constraints, ensuring the least possible deviation from the unconstrained trajectory.

Gauss' principle of least constraint can be applied to any dynamical system including the two-dimensional motion of incompressible fluids described by the 2D incompressible NSE, which can be written in the following form:
\begin{align}
    \frac{\partial \mathbf{u}}{\partial t} + (\mathbf{u} \cdot \nabla)\mathbf{u} &= -\nabla p + \nu \nabla^2\mathbf{u}, \label{NSEa} \\
    \nabla \cdot \mathbf{u} &= 0, \label{NSEb}
\end{align}
where $\mathbf{u} = (u, v)^T$, is the fluid velocity vector, $p$ is the pressure, and $\nu$ is the kinematic viscosity.  Equation (\ref{NSEa}) represents a balance of forces applied on the moving fluid in a domain $\Omega \in R^2$. The left-hand side of the Equation is the total acceleration of the fluid, incorporating both inertial and convective components, while the right-hand side comprises the forces acting on the fluid, which can be decomposed into either external or constraint forces.  Finally, Equation (\ref{NSEb}) represents a mass balance often referred to as continuity. 

\begin{figure}[h!]
\centering
\includegraphics[width=0.4\textwidth]{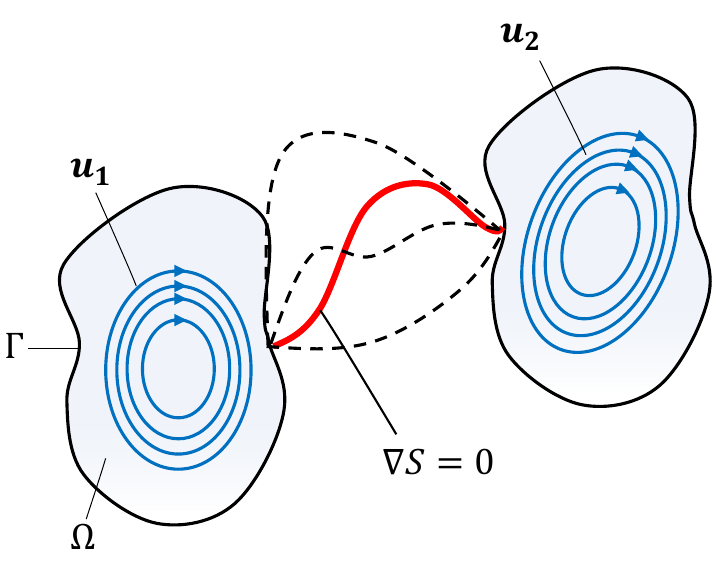}
\caption{Illustration of the evolution of the flow field in the space-time configuration, highlighting the path traced by $\mathbf{u}$ with a stationary quantity $\nabla S = 0$.}
\label{fig:flowSchematic}.
\end{figure}

Several works \citep{gresho1987pressure, gonzalez2022variational} have demonstrated that in incompressible flow, the pressure gradient function acts as a constraint force, primarily enforcing the continuity. Hence, by applying Gauss' principle of least constraint to the dynamics of incompressible fluids governed by NSE, and classifying the pressure gradient as a constraint force, we can write the action, $S$, in Euler coordinates as:
\begin{equation}
    S = \displaystyle \frac{1}{2} \int_{\Omega} \rho \left(\frac{\partial \mathbf{u}}{\partial t} + \mathbf{u} + \nabla \mathbf{u} - \nu \Delta \mathbf{u} \right)^2 d\mathbf{x}, \label{actionS}
\end{equation}
subject to the continuity 
\begin{equation}
    \nabla\; . \; \mathbf{u} = 0,  
    \label{actionConstraint}
\end{equation}
and any boundary conditions defined on the boundary of the domain $\Gamma = \partial \Omega$.

As shown in Fig. \ref{fig:flowSchematic}, when the system evolves, multiple trajectories of the flow field $\mathbf{u}$ become possible, of which only few are shown. The trajectory followed by the system, represented by the solid line, is the one that renders $\nabla S = 0$. Thus, the flow field will deviate from the motion dictated by the inertia and the viscous forces only by the amount necessary to satisfy the continuity. In other words, no larger pressure gradient will be generated than necessary to satisfy the continuity \citep{taha2023minimization}. Thus, the name the principle of minimum pressure gradient (PMPG).

\section{Physics-Informed Learning Using Navier Stokes Equation}
In general, PINN integrates information from the governing physical laws of a given system into the training process of a deep neural network (NN) so that the unknown variables can be approximated using a limited set of training samples. In the case of the NSE, Equations \eqref{NSEa}, \eqref{NSEb}, the unknown flow field variables, $u$, $v$, as well as the pressure, $p$, are attained by converting the process of finding the solution of the PDEs into an optimization problem in which a loss function (error function) is iteratively minimized by updating the parameters of the neural network. This process is denoted by NN training.

\begin{figure*}[t!]
\centering
\includegraphics[width=\textwidth]{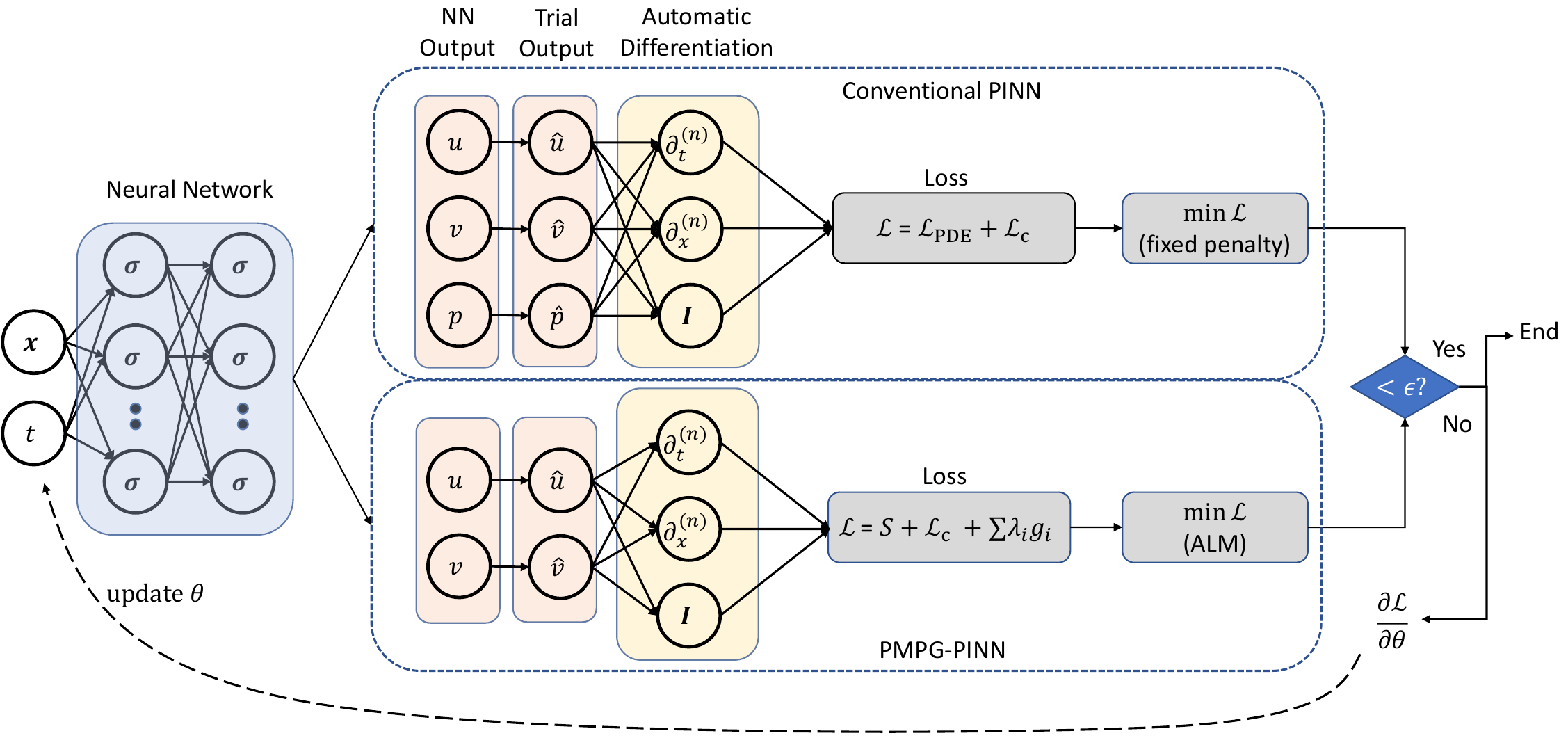}   
\caption{Schematic of a physics-informed neural network showing both conventional PINN, and PMPG-PINN schemes.}\label{PINNSchematic}.
\end{figure*}

As shown in Fig. \ref{PINNSchematic}, the NN is a fully-connected feed-forward network composed of $L$ multiple-hidden layers. It takes a concatenation of time and the state-space $z^0 = (t, \mathbf{x})$ as an input, and outputs a guess for the unknown variables ($u,v,p$).  Each layer creates data for the next layer through a tensorial nested transformation of the form \citep{svozil1997introduction}:
\begin{equation}
 z^l = \sigma^l (W^l.z^{l-1} + b^l), \quad l = 1, 2, \ldots, L, 
\end{equation}
where the functions $\sigma^l$ are called activation functions. These can be chosen based on nature of the problem to be solved.  The variables $W^l$ and $b^l$ denote, respectively, the weights and biases of each NN layer, $l$. These are updated after every iteration (epoch) by minimizing a loss function, $\mathcal{L}$, with respect to those variables. The loss function, denoted here by $\mathcal{L}(\theta)$ where $\theta=[W^l,b^l]$, measures the difference between the predicted solution and the target solution. When the the residual change between successive iterations is less than a predefined threshold, say $\epsilon$, the training stops and the output of the last iteration is considered the solution of the PDE. Further details of this process are presented in Appendix A. 

In the case of Equation \eqref{NSEa}, \eqref{NSEb}, the loss function, $\mathcal{L}(\theta)$, is defined as: 
\begin{equation}
    \mathcal{L}(\theta)  = \frac{\mu_{p}}{2}  \mathcal{L}_{PDE} + \frac{\mu_c}{2}  \mathcal{L}_c,
     \label{loss_NSE}
\end{equation}
and is formed by adding two losses subject to predefined weights (penalties). The first is the loss function, $\mathcal{L}_{PDE}$, weighted by $\mu_p$, which defines the loss in the calculation of Equation \eqref{NSEa} and is given by:
\begin{equation}
    \mathcal{L}_{PDE} = \frac{1}{N_B} \sum_{i}^{N_B} f_i(t, \mathbf{x})^2, \quad f =   \frac{\partial \mathbf{u}}{\partial t} + (\mathbf{u} \cdot \nabla)\mathbf{u} 
   +\nabla p - \nu \nabla^2\mathbf{u}.
\end{equation}
where $N_B$ is the number of collocation points over the domain. The second is the loss function, $\mathcal{L}_{c}$, weighted by $\mu_c$, which defines the loss in the calculation of the continuity, Equation \eqref{NSEb}, and is defined as: 
\begin{equation}
    \mathcal{L}_{c} = \frac{1}{N_B} \sum_{i}^{N_B}  g_i(\mathbf{x})^2, \quad g = \nabla\,. \mathbf{u}.
\end{equation}
Note that, the mean square error (MSE) is employed to compute the losses. 

\section{PMPG-Based Physics-Informed Learning}
Deviating from the conventional approach of including the loss function associated with NSE into the training process, the proposed PMPG-PINN searches for the velocity field $\mathbf{u}$ which minimizes the function $S$ in Equation \eqref{actionS} subject to the continuity constraint \eqref{actionConstraint} and the associated boundary conditions. Thus, the pressure and its gradients are completely eliminated from the training process. This can be simply achieved by converting the constrained optimization problem into an unconstrained optimization problem using the following loss function:
\begin{equation}
    \mathcal{L}(\theta)  = S + \frac{\mu_c}{2}  \mathcal{L}_c,
    \label{loss_PMPG}
\end{equation}
where the integral $S$ can be reformulated as a sum using the mean rule, leading to
\begin{equation}
    S = \frac{A}{N_B} \sum_{i}^{N_B}     J_i,
\end{equation}
Here $A$ is the area of the domain, and $J$ is the discretized operator used to approximate the integrand in Equation \eqref{actionS}, and can be written as: 
\begin{equation}
    J = \displaystyle \frac{1}{2}\rho \left(\frac{\partial \mathbf{u}}{\partial t} + \mathbf{u} + \nabla \mathbf{u} - \nu \Delta \mathbf{u} \right)^2.
\end{equation}

In comparing the definitions of the loss functions in Equation \eqref{loss_NSE} and Equation \eqref{loss_PMPG}, a key distinction emerges. The optimization problem in Equation \eqref{loss_NSE} is easier to solve given that the losses associated with the PDE and the continuity, $\mathcal{L}_{PDE}$ and $\mathcal{L}_c$, can be jointly driven to zero. This consistency enables simultaneous achievement of both constraints. In contrast, for the PMPG-PINN problem, the minimization of the loss function is achieved even when the objective function, $S$, is different from zero. Thus, a scheme that merely optimizes the sum of the objective and the constraint loss will usually end up in an optimum that is not a solution of the PDE \citep{nocedal2006penalty}. 

To overcome this issue, we utilize the Automated Lagrangian Method (ALM) \citep{nocedal2006penalty}, which replaces the constrained optimization problem with a sequence of unconstrained problems with an additional term designed to mimic Lagrangian multipliers. As such, the updated loss within this methodology at the k-th iteration can be written as: 
\begin{equation}
    \mathcal{L}(\theta)  = S + \frac{\mu_c}{2N_B} \sum_{i}^{N_B} g_i(\mathbf{x})^2 - \frac{1}{N_B}\sum_{i}^{N_B} \lambda_i^k g_i(\mathbf{x}),
\end{equation}
where $\mu_c$ is the penalty weight, and $\lambda_i^k$ are the Lagrange multipliers. Additional details specifying the algorithm governing the updates of $\lambda_i^k$ following each iteration can be found in Appendix B. 

\section{Example: Lid-Driven Cavity}
To illustrate the effectiveness of the proposed approach, we consider the widely-known lid-driven cavity problem. Among the many reasons which made this particular problem one of the most common benchmarks in CFD is the combination of its simple geometry and the presence of various corner singularities. Furthermore, only velocity Dirichlet boundary conditions are required to define the mathematical problem. We initially tackle the problem using conventional PINN. This initial step sets a baseline, enabling a comparison with the new approach. Following this, we proceed to illustrate the application of the PMPG-PINN method in two scenarios; one while employing Lagrange multipliers  and the other without. Finally, we evaluate the computational efficiency of PMPG-PINN by comparing it to the conventional PINN. 

\begin{figure}
\centering
\includegraphics[width=0.4\textwidth]{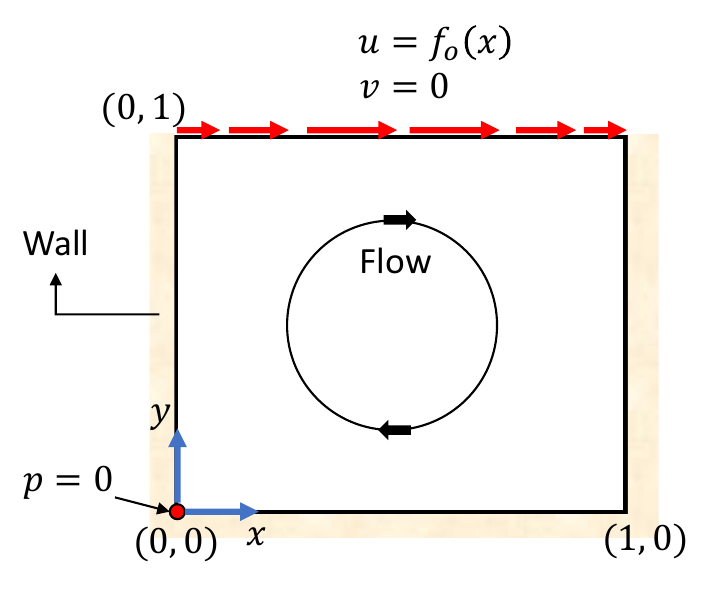}
\caption{Schematic diagram of the lid-driven cavity problem showing the domain and boundary conditions.}\label{schematic}.
\end{figure}

As shown in Fig. \ref{schematic}, the computational domain is defined on the area [0, 1] x [0, 1].  A no-slip condition is enforced on three walls, while the upper wall is assumed to move to the right with a velocity $u = f_0(x)$, where, in order to ensure continuous differentiability along the boundaries, $f_0(x)$ is assumed to be a parabolic function in $x$. Furthermore, zero-pressure is prescribed at the lower left cavity corner.  The maximum axial velocity is chosen such that the flow remains laminar with a Reynolds number that is equal to Re = 100.

Since the problem considered is stationary, the time variable is eliminated and the traditional PINN takes $\mathbf{x} = (x, y)$ as inputs, and outputs the velocity and pressure fields as ($u, v, p$). To force the output solution to satisfy the boundary conditions, we modify the network and construct a new trial solution, ($\hat{u}, \hat{v}, \hat{p}$), as:  
 \begin{align}
 \hat{u} &= yf_0(x) + x(1-x)y(1-y)u,\\
 \hat{v} &=  x(1-x)y(1-y)v,\\
 \hat{p} &= (x + y) p. 
 \end{align}
 
During the training process utilizing the conventional PINN, the PDE and continuity were equally penalized by using equal loss weights ($\mu_p=\mu_c=2$).  A parametric study was then performed to determine the minimum number of neurons per layer, $n$, collocation points, $N_B$, and network layers, $L$, that minimize computational time while leading to acceptable residual errors. This led to the following numerical values: $n = 20$, $L = 8$ and $N_B = 90000$, which were used throughout the presented simulations.

\begin{figure}
\centering
\includegraphics[width=0.45\textwidth]{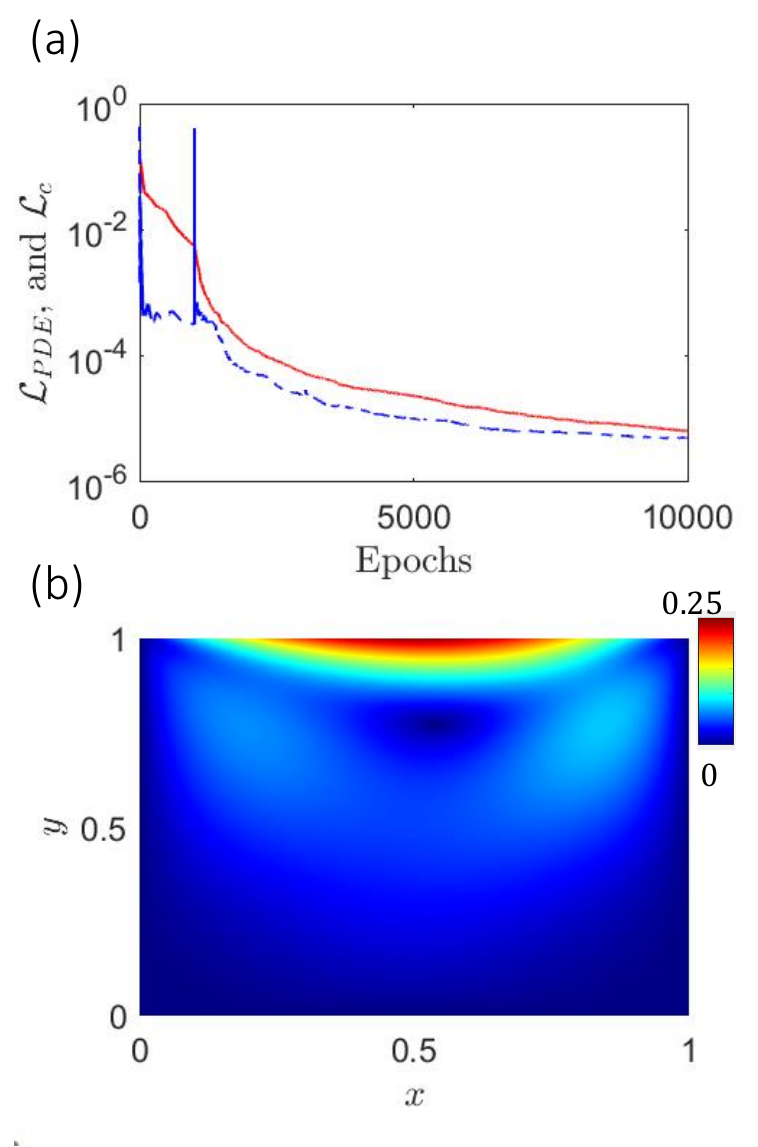}
\caption{A physics-informed solution of the Lid-driven cavity problem using conventional PINN. (a) Convergence of the residual loss functions, (dashed) $\mathcal{L}_{PDE}$, and (line) $\mathcal{L}_c$. (b) Heat map of the velocity magnitude: $\sqrt{u^2+v^2}$.}\label{fig:PINN_NSE}.
\end{figure}

Figure \ref{fig:PINN_NSE}a presents convergence of the residual loss functions, $\mathcal{L}_{PDE}$, and $\mathcal{L}_{c}$, with the number of epochs as obtained using the conventional PINN. Both residuals demonstrate a consistent decrease until they reach an acceptable value in the order of $10^{-6}$ after nearly 10000 epochs. The number of epochs differs based on the initialization of the PINN, yet the computational time per epoch remains nearly constant at \SI{0.050}{\second}. The velocity magnitude resulting from these computations is depicted in Fig. \ref{fig:PINN_NSE}(b). 

\begin{figure}
\centering
\includegraphics[width=0.4\textwidth]{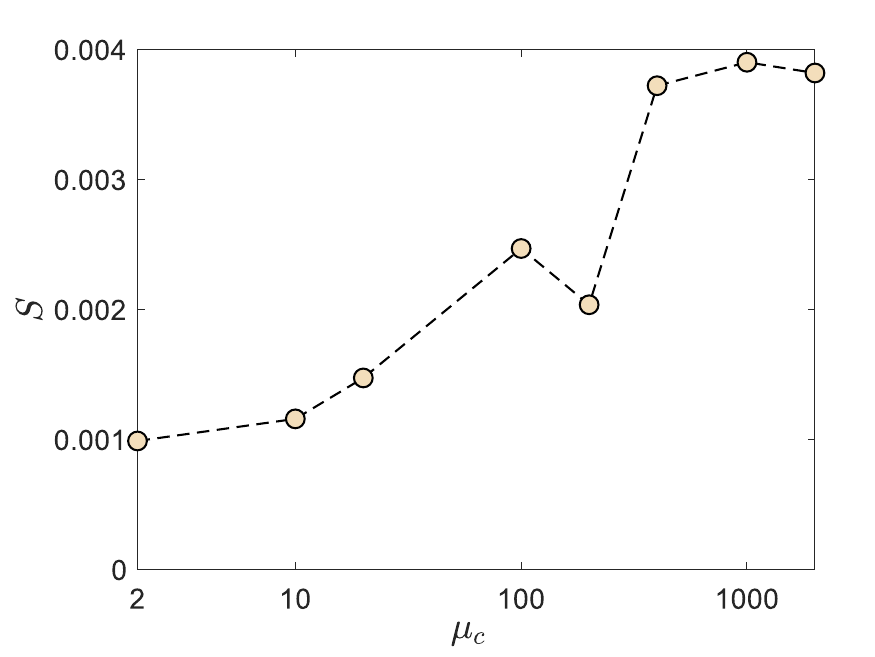}
\caption{Variation of the quantity $S$ with the penalty weight $\mu_c$.}
\label{fig:penalty1}
\end{figure}

Next, PMPG-PINN is implemented on the same domain with the same network parameters. The first implementation does not use Lagrange multipliers in order to determine if the fixed weight scheme can result in accurate results. The value of $S$ which minimizes the loss function is obtained for fixed values of $\mu_c$ ranging between 2 and 2000.  Figure \ref{fig:penalty1} demonstrates that $S$ varies considerably with $\mu_c$, which indicates that the fixed-weight training scheme results in erroneous predictions that are weight dependent.  To see how far those values of $S$ are from the reference value that the conventional PINN converges to, we substitute the velocity field obtained using the conventional PINN into Equation \eqref{actionS} and obtain a reference value of $S = 0.002$. It is apparent that, because the continuity constraint is not highly penalized when $\mu_c$ is small, the Langrange-less PMPG-PINN yields a value of $S$ that is much lower than the reference value. A higher penalty of $\mu_c= 200$ yields the best enforcement of the continuity constraint and a value of $S\approx0.002$. However, contrary to natural intuition, penalizing the continuity further causes the value of $S$ to diverge from the reference value because the optimization problem becomes ill-conditioned \cite{nocedal2006penalty}. 

\begin{figure}
\centering
    \includegraphics[width=0.45\textwidth]{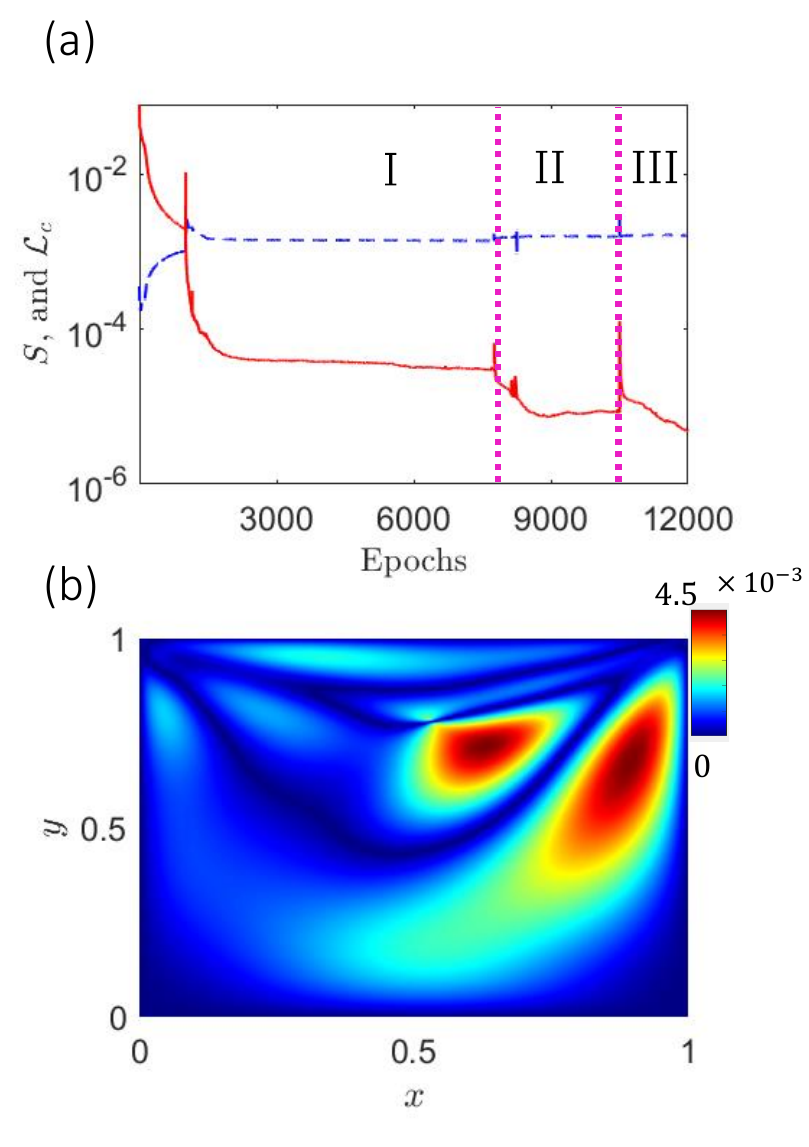}
\caption{A physics-informed solution of the Lid-driven cavity problem using PMPG-PINN. (a) Convergence of the residual loss functions for $\mu_c = 10$, (dashed) $S$ and (line) $\mathcal{L}_c$. (b) Point-wise error of the velocity magnitude between conventional PINN and PMPG-PINN with $\mu_c = 50$. }\label{PINN_Lag}.
\end{figure}

Next, we explore the benefits of using Lagrange multipliers in the loss function using two different penalties; namely $\mu_c=10$ and $\mu_c=50$. For the case $\mu_c = 10$, both the value of $S$ and the residual of the loss constraint, $\mathcal{L}_c$, are plotted against the number of epochs in Fig. \ref{PINN_Lag}(a). The figure is sectioned into three areas, each representing one outer iteration in which the Lagrange multiplier is updated. It can be seen that adjustments to the Lagrange multipliers after every outer iteration following the procedure described in Appendix \ref{appendALM} correlates with significant reductions in the value of the residual yielding a value of $S = 0.0018$ and a residual of the order of $10^{-6}$ after nearly 12000 epochs. 

When PMPG-PINN is trained using $\mu_c = 50$, the value of $S$ converges to $0.00195$, while the residual of the constraint loss, $\mathcal{L}_c$, decreases to an acceptable value of the order of $10^{-6}$ after nearly 11000 epochs. Figure \ref{PINN_Lag}(b) illustrates the point-wise error of the velocity magnitude between conventional PINN presented earlier and PMPG-PINN for the case of $\mu = 50$. It can be clearly seen that the error is low, implying a good agreement between the two frameworks. In general, the accuracy of the PMPG-PINN framework can be improved further by considering higher-order integration methods of $S$, such as trapezoidal or Simpson's rule \citep{kreider2005principles}. Beyond this, several other suggestions could be considered, such as, network architecture tuning, and learning rate optimization.

When implementing PMPG-PINN, we observed a notable enhancement in the computational efficiency. Specifically, the time required per epoch in PMPG-PINN is measured at \SI{0.044}{\second}, in contrast to \SI{0.050}{\second} observed using the conventional PINN. This 12\% reduction in training time per epoch was found to be consistent regardless of the value of $\mu_c$, the initial guesses of the NN parameters, or whether ALM is implemented or not. The reduction in time can be attributed to two factors: $i)$ reduction in the number of automatic derivative calculations due to the elimination of the pressure gradients, and $ii)$ reduction in the output variables, which means fewer parameters to adjust and consequently a smaller output space to learn. The reduction in time per epoch when implementing PMPG-PINN highlights its potential applicability in solving complex and computationally demanding problems.  It is important to note, however, that this reduction in computational cost comes with a trade-off because direct calculation of the pressure field is not part of PMPG-PINN framework. Nevertheless, the pressure field can be recovered by applying the resultant velocity field, $u$ and $v$, to Equation \ref{NSEa}, followed by integrating the resulting linear, first-order, and time-independent PDE.

\section{Conclusions}
In this paper, we developed a new approach to solve incompressible flow problems by integrating the Principle of Minimum Pressure Gradient (PMPG) with Physics-Informed Neural Networks (PINN) to create the PMPG-PINN framework. A critical component in the development of the new approach is the incorporation of the Augmented Lagrange Multipliers (ALM) in the loss function of the neural network. This implementation is key towards effectively balancing the objective and the constraint of the model.

Applying the PMPG-PINN framework to the stationary lid-driven cavity problem yielded results that are in excellent agreement with those obtained using the conventional PINN approach.  Moreover, PMPG-PINN results in nearly 12\% reduction in computational time per epoch during neural network training independent of the network initialization and the penalty imposed on the loss function. The enhanced efficiency is primarily attributed to the elimination of pressure gradient calculations and the reduction of output parameter space in the neural network. 

We believe that PMPG-PINN may offer a potentially important alternative to the conventional PINN based on NSE. Its apparent computational efficiency can pave the way toward solving  more complex and non-stationary problems. For example, a natural progression would be to adapt the PMPG-PINN framework to transient and turbulent flow conditions.  Venturing into the domain of fluid-structure interaction presents another frontier for exploration. Here, a key challenge lies in integrating the dynamics of solids into the formulation of the quantity $S$, which in turn can be minimized using PMPG-PINN framework. 

\section*{Data Availability Statement}
The data that support the findings of
this study are available from the
corresponding author upon reasonable
request

\appendix 
\section{NN structure, activation and optimizer}
A fully-connected neural network with $L$ hidden layers and $n$ neurons per layer is employed. Because of its effectiveness, we used the \textit{tanh} activation function \citep{nwankpa2018activation}. The weights of the network are initialized with Xavier initialization \citep{glorot2010understanding},  while the biases are initialized as zero. No dropout methods were applied. To minimize the loss function,  $\mathcal{L}$, we combine ADAM \citep{kingma2014adam} and L-BFGS-B\citep{zhu1997algorithm} optimizers. We first apply the ADAM optimizer for gradient descent training and then employ the L-BFGS-B optimizer to fine tune the results. During the Adam-based training, the optimizer calculates the direction of the gradient at each iteration considering a full-batch size. The initial learning rate is 0.001. We set the stopping criterion of L-BFGS-B to the smallest positive normalized floating-point number represented in Python3. 

\section{Augmented Lagrangian Method (ALM)}\label{appendALM}
The Augmented Lagrangian Method (ALM) is an optimization technique designed to handle constraints more effectively in optimization problems. It is particularly useful in scenarios where conventional methods might suffer from ill-conditioning issues \citep{nocedal2006penalty}.  The ALM modifies the loss function defined in Equation \eqref{loss_PMPG} by incorporating explicit Lagrange multiplier estimates, $\lambda_i^k$, which transforms the original constrained optimization problem into a series of unconstrained problems, each characterized by varying coefficients of $\lambda_i$. Thus, the new ALM loss function can be written as:  
\begin{equation}
    \mathcal{L}(\theta)  = S + \frac{\mu_c}{2N_B} \sum_{i}^{N_B} g_i(\mathbf{x})^2 - \frac{1}{N_B}\sum_{i}^{N_B} \lambda_i^k g_i(\mathbf{x}).
\end{equation}

This ALM process involves adjusting the Lagrange multipliers to ensure that the constraints are increasingly satisfied as the outer iterations proceed. During each outer iteration, the PINN algorithm seeks an approximate minimizer $\theta_k$, which will then become $\theta_{k+1}$. As such, we can write the optimality condition as:
\begin{align}
        0 =& \nabla_\theta \mathcal{L}(\theta_{k+1}) = \nabla S(\theta_{k+1})\, - \nonumber \\&\frac{1}{N_B}\sum_{i}^{N_B} [\lambda_i^{k} - \mu_c g_i(\mathbf{x};\theta_{k+1})]  \nabla g_i(\mathbf{x}; \theta_{k+1}) 
\end{align}
thus, 
\begin{equation}
   \nabla S(\theta_{k+1}) = \frac{1}{N_B}\sum_{i}^{N_B} [\lambda_i^{k} - \mu_c g_i(\mathbf{x}; \theta_{k+1})] \nabla g_i(\mathbf{x}; \theta_{k+1}) \label{alg1}
\end{equation}
In the $k+1$ iteration, we anticipate that the constraint function
$g$ is nearly satisfied, and consequently rendering $g^2$ close to zero. Therefore, the Lagrange multipliers, $\lambda_i^{k+1}$, must be updated to replicate the gradient in the previous iteration as:
\begin{equation}
        \nabla S(\theta_{k+1}) = \frac{1}{N_B} \sum_{i}^{N_B} \lambda_i^{k+1} \nabla g_i(\mathbf{x}; \theta_{k+1}) \label{alg2}
\end{equation}
Comparison of \eqref{alg1} and \eqref{alg2} indicates the correct update formula for $\lambda^{k+1}$ is:
\begin{equation}
    \lambda_i^{k+1} = \lambda_i^{k} - \mu_c g_i(\mathbf{x};\theta_{k+1}). \label{updateLambda}
\end{equation}

A general framework for algorithm based on the ALM can be specified as follows:
\begin{algorithm}
\SetAlgoLined
\DontPrintSemicolon
\KwData{: Initialize $\mu_c$. Set $\lambda^0 = 0$.}
\For{$k = 0, 1, 2, \ldots$}{
    Starting at $\theta_k$, find an approximate minimizer of $\mathcal{L}(;\mu_c, \lambda_i^k)$\;
        \If{$\nabla_\theta \mathcal{L}(\theta) \leq \epsilon$}{

        stop with approximate solution $\theta_{k+1}$\;
    }
    Choose new $\lambda^{k+1}$ based on Equation \eqref{updateLambda}\;
    Choose new starting point $\theta_{k+1}$\;
}
\caption{ALM framework}
\end{algorithm}

\nocite{*}

\bibliography{aipsamp}
\end{document}